\documentclass[12pt]{elsart}
\usepackage{amssymb}
\usepackage{ecltree}
\usepackage{latexsym}
\usepackage{amsfonts}
\usepackage{epsfig}
\usepackage{pifont}
\begin{document}
\def\bfone{\relax{\rm 1\kern-.35em 1}}
\def\bfnull{\relax{\rm O \kern-.635em 0}}
\def\dop{{\rm d}\hskip -1pt}
\def\mez{\frac{1}{2}}
\def\sx{\left}
\def\dx{\right}
\def\na{\nabla}
\def\imez{\frac{i}{2}}
\def\a{\alpha}
\def\b{\beta}
\def\g{\gamma}
\def\d{\delta}
\def\e{\epsilon}
\def\ve{\varepsilon}
\def\t{\theta}
\def\l{\lambda}
\def\m{\mu}
\def\n{\nu}
\def\pg{\pi}
\def\r{\rho}
\def\s{\sigma}
\def\t{\tau}
\def\z{\zeta}
\def\c{\chi}
\def\p{\psi}
\def\o{\omega}
\def\G{\Gamma}
\def\D{\Delta}
\def\T{\Theta}
\def\L{\Lambda}
\def\Pg{\Pi}
\def\S{\Sigma}
\def\O{\Omega}
\def\pb{\bar{\psi}}
\def\cb{\bar{\chi}}
\def\lb{\bar{\lambda}}
\def\i{\imath}
\def\Pii{\mathcal{P}}
\def\Q{\mathcal{Q}}
\def\K{\mathcal{K}}
\def\A{\mathcal{A}}
\def\N{\mathcal{N}}
\def\F{\mathcal{F}}
\def\Gi{\mathcal{G}}
\def\Ci{\mathcal{C}}
\def\oL{\overline{L}}
\def\eq#1{(\ref{#1})}
\newcommand{\be}{\begin{equation}}
\newcommand{\ee}{\end{equation}}
\newcommand{\ba}{\begin{eqnarray}}
\newcommand{\ea}{\end{eqnarray}}
\newcommand{\ban}{\begin{eqnarray*}}
\newcommand{\ean}{\end{eqnarray*}}
\newcommand{\nn}{\nonumber}
\newcommand{\nin}{\noindent}
\newcommand{\fgl}{\mathfrak{gl}}
\newcommand{\fu}{\mathfrak{u}}
\newcommand{\fsl}{\mathfrak{sl}}
\newcommand{\fsp}{\mathfrak{sp}}
\newcommand{\fusp}{\mathfrak{usp}}
\newcommand{\fsu}{\mathfrak{su}}
\newcommand{\fp}{\mathfrak{p}}
\newcommand{\fso}{\mathfrak{so}}
\newcommand{\fg}{\mathfrak{g}}
\newcommand{\fr}{\mathfrak{r}}
\newcommand{\fe}{\mathfrak{e}}
\newcommand{\rE}{\mathrm{E}}
\newcommand{\rSp}{\mathrm{Sp}}
\newcommand{\rSO}{\mathrm{SO}}
\newcommand{\rSL}{\mathrm{SL}}
\newcommand{\rSU}{\mathrm{SU}}
\newcommand{\rUSp}{\mathrm{USp}}
\newcommand{\rU}{\mathrm{U}}
\newcommand{\rF}{\mathrm{F}}
\newcommand{\R}{\mathbb{R}}
\newcommand{\C}{\mathbb{C}}
\newcommand{\Z}{\mathbb{Z}}
\newcommand{\Hb}{\mathbb{H}}
\def\oL{\overline{L}}

\begin{frontmatter}
\title{On $D=4$, $N=2$ Supergravity with Abelian electric and magnetic charges}
\author[luca]{Luca Sommovigo} and \author[silvia]{Silvia Vaul\`a}
\address[luca]{Dipartimento di Fisica, Politecnico di Torino Corso Duca degli Abruzzi 24, I-10129 Torino, Italy and Istituto Nazionale di Fisica Nucleare (INFN) - Sezione di Torino, Via P. Giuria 1, I-10125 Torino, Italy}
\address[silvia]{DESY, Theory Group Notkestrasse 85, Lab. 2a D-22603 Hamburg, Germany\\ and \\
II. Institut f\"ur Theoretische Physik der Universit\"at Hamburg,\\
 Luruper Chausse 179, D-22761 Hamburg, Germany}
\ead{luca.sommovigo@polito.it} \ead{silvia.vaula@desy.de} 
\begin{abstract}
 We discuss the relation between standard $N=2$ supergravity with
 translational gauging and $N=2$ supergravities with scalar--tensor
 multiplets with massive tensors and Abelian electric charges. We
 point out that a symplectic covariant formulation of $N=2$
 supergravity can be achieved just in the presence of tensor
 multiplets. As a consequence one can see that the formulation of the
 $N=2$ theory as it comes from IIB flux compactification, which is
 included in these models, is equivalent to a non perturbative phase
 of standard $N=2$ supergravity. It is also shown that the IIB tadpole
 cancellation condition is imposed by supersymmetry in four
 dimensions.   
\end{abstract}

\begin{keyword} Supergravity, Flux Compactification
\PACS 04.65.+e, 11.25.Mj
\end{keyword} 

\end{frontmatter}

\section{Introduction}
In compactification from IIB supergravity, a double tensor multiplet
naturally arises from the zero modes of the RR and NSNS 2--forms and
from the complex dilaton \cite{Berkovits:1995cb}. In absence of fluxes
the two massless tensors can be dualized into scalars and one obtains
the standard $N=2$ supergravity coupled to vector multiplets and
hypermultiplets \cite{deWit:1982na}. In the presence of fluxes the
relation with standard gauged supergravity becomes less trivial,
particularly when magnetic fluxes are included.\\ 
In ref. \cite{Dall'Agata:2003yr}, the $N=2$ gauged supergravity theory
coupled to $n_V$ vector multiplets and $n_T$ scalar--tensor multiplets
was analyzed, extending a previous work \cite{Theis:2003jj} where the 
$N=2$ Lagrangian of ungauged supergravity coupled to $n_T$
scalar--tensor multiplet  in the absence of vector multiplets was
constructed.\\ 
Besides the terms due to the semisimple gauging and to the masses of
the tensors, as discussed in ref. \cite{Dall'Agata:2003yr}, the
fermionic shifts and the scalar potential may also contain terms
involving Abelian electric charges \cite{Taylor:1999ii},
\cite{Louis:2002ny} which together with the mass terms reconstruct
symplectic invariant structures which are the remnant of the IIB
electric and magnetic fluxes. The geometrical interpretation of both
Abelian electric charges and the mass parameters, becomes more clear
in terms of the standard $N=2$ supergravity where the Abelian
isometries, associated to the axions which are dualized into tensors,
are gauged before the dualization procedure.\\ 
In fact if the ordinary gauging is performed after dualization of the
axions into tensors, the reduced scalar manifold does not have any
translational symmetry left for the residual quaternionic coordinates,
therefore the Abelian electric charges associated to the axionic
symmetries cannot appear in the potential. If however the
translational gauging is performed before the dualization procedure,
then the Abelian electric charges are present from the very beginning
in the fermionic shifts, and hence in the potential, as shown in
Section 2. \\
By means of a symplectic rotation, one can generate the mass terms for
the tensors , which can be interpreted as magnetic charges
\cite{Louis:2002ny}.  Anyway, such a symplectic rotation acts non
perturbatively on the theory, as it is evident on the gauged
supergravity side.\\  
For this purpose, in Section 3, we act with a symplectic rotation on
the Bianchi identities/equations of motion of the standard gauged 
supergravity. In this way one gets also a source for the magnetic
field strengths and a formulation in terms of a Lagrangian may be
problematic. However by a judicious choice of the symplectic
transformations, together with the dualization of the axions into
tensors, one can still find a Lagrangian.\\ 
This dualization procedure clearly shows that the $N=2$ supergravity
which gives a symplectic invariant potential, as it comes from IIB
compactifications on a Calabi--Yau in the presence of electric and
magnetic fluxes \cite{Taylor:1999ii,Louis:2002ny}, must be a theory
with scalar--tensor multiplets.\\  
We also show that the supersymmetry Ward identity for the scalar
potential, implies:$$e_\L^I m^{\L J}- e_\L^J m^{\Lambda  I}=0$$ 
where $e_\Lambda^I, m^{\Lambda J}$ are the electric and magnetic
charges, playing the r\^ole of electric and magnetic fluxes
respectively. This corresponds to the ten dimensional tadpole
cancellation condition (when we restrict $I=1,2$),  hence
supersymmetry implies that we are in the local case, so that with a
{\it non perturbative} symplectic rotation we can always reduce to the
case with only electric charges present. Let us stress, however, that
the Lagrangian is not  invariant under non perturbative
transformations, therefore such a symplectic rotation is not allowed
at the level of the low energy Lagrangian. This means that the
Lagrangian which descends from IIB flux compactification, when also
magnetic fluxes are present, is not equivalent to the Lagrangian of
standard $N=2$ supergravity with translational gauging. The
identification holds just at the perturbative level where the
parameters $m^{I\L}$ are zero.  

\section{The dualization with translational gauging}
Let us consider the terms in the  $N=2$ bosonic Lagrangian with
translational gauging, involving the hypermultiplets scalars and the
vector fields \footnote{ For all notations and conventions we refer to
  \cite{Dall'Agata:2003yr}, \cite{Andrianopoli:1996cm}}: 
\be\label{boslagG}
\mathcal{L}=\frac{1}{2}\F^\L \Gi_\L
-h_{uv} dq^{u}*dq^{v}-2h_{Iu}\nabla q^I*dq^u-h_{IJ}\nabla q^I*\nabla
q^J-V*\bfone 
\ee We have split the quaternionic coordinates in the following way:
$q^{\hat u}=(q^u,q^I)$, $\hat u=1,\dots 4n$, $I=1,\dots n_T$, $
u=n_T+1,\dots 4n $, where $n$ is the number of hypermultiplets and
$n_T$ is the number of translational isometries on the quaternionic
manifold. We chose the coordinates on the quaternionic manifold in
such a way that the translations are realized as $q^I\rightarrow
q^I+\eta^I$, therefore the Lagrangian does not depends on the bare
fields $q^I$.\\ 
The electric and magnetic field strengths are defined as usual:
\be \F^\L=dA^\L;\quad
\Gi_\L\equiv\frac{\d\mathcal{L}}{\delta\F^\L}=-Im\N_{\L\S} *\F^\S
+Re\N_{\L\S}\F^\S\ee 
We will focus our attention on the quaternionic manifolds which enjoy
the following property: 
\be\label{quatquat}\o^x_I\o^x_J=h_{IJ}\ee
which certainly holds for Type IIB theory on Calabi--Yau
\cite{Dall'Agata:2001zh}, \cite{Louis:2002ny}, even thougth
the whole procedure can be applied to the general case. When equation
\eq{quatquat} holds the scalar potential is given by:
\be\label{pot1}
V=-\frac{1}{2}\Pii^x_\L(Im\N^{-1})^{\L\S}\Pii^x_\S;\quad
\Pii^x_\L(q^u)\equiv e^I_\L\o^{x}_I(q^u)\ee 
where $\Pii^x_\L$ are the quaternionic prepotentials. The covariant
derivative of the axions is defined as: 
\be \nabla q^I=dq^I+\A^\L e_\L^I\label{decu}\ee
where $e_\L^I$ are constant killing vectors which are $(m+1)\times
n_T$ rectangular matrixes which select which combination of the $m+1$
vectors gauges each translational isometry. Since these are $n_T$
linearly independent combinations, by means of a block diagonal
symplectic matrix $\bf{M}$ 
\be{\bf M}=\begin{pmatrix}{\mathcal{M}&0\cr
    0&\mathcal{M}}\end{pmatrix};\quad
\mathcal{M}^T\mathcal{M}=\bfone\ee 
we can go into a basis where each of these combinations coincides with
a vector $\A^{\dot{\L}}$, with $\dot{\L}=0\dots n_T$, in such a way
that the rectangular matrix $e_\L^I$ contains as a submatrix the
$n_T\times n_T$ matrix $e_{\dot{\L}}^I$ while all the other entries
are zero, that is $e^I_{\tilde{\L}}=0$ with $\tilde{\L}=n_T+1\dots
m+1$.  
The equations of motion of the vector fields, together with the
Bianchi identities, in this particular basis are given by: 
\be\label{bieom}
d\begin{pmatrix}{\F^{\dot\L}\cr\F^{\tilde\L}\cr\Gi_{\dot\L}\cr\Gi_{\tilde\L}}
\end{pmatrix}=\begin{pmatrix}{0\cr0\cr 
    2e^I_{\dot\L}\left(h_{Iu}*dq^u+h_{IJ}*\nabla q^J\right)\cr
    0}\end{pmatrix}\equiv\begin{pmatrix}{0\cr 0\cr e^I_{\dot\L}J_I\cr
    0}\end{pmatrix}\ee 
Let us consider the equations of motion of the hypermultiplet scalars:
\ba&&\label{eomqI} \frac{\d\mathcal{L}}{\d q^I}=0\Rightarrow
    d(2h_{Iu}*dq^u+2h_{IJ}*\nabla q^J)=0\Leftrightarrow dJ_I=0\\[2mm] 
&&\label{eomqu}\frac{\d\mathcal{L}}{\d q^u}=0\Rightarrow\
    d(2h_{uv}*dq^v+2h_{Iu}*\nabla q^I)-\frac{\partial V}{\partial
    q^u}+\nn\\[2mm] 
&&-\partial_u h_{wv}dq^w*dq^v-2\partial_u h_{Iv}\nabla
    q^I*dq^u-\partial_u h_{IJ}\nabla q^I*\nabla q^J=0 
\ea
One can easily see that equation \eq{eomqI} is solved if we set 
\be\label{sburz}h_{Iu}*dq^u+h_{IJ}*\nabla q^J=dB_I\equiv H_I\ee
where $B_I$ is an arbitrary 2--form, such that the equation of motion
    for the scalars corresponds to the Bianchi identity of the
    2--form: 
\be d^2B_I=0\ee
Applying the Hodge duality to equation \eq{sburz} and further
    differentiating, we obtain the
    equation of motion for the $B_I$'s from the Bianchi identities of
    the $q^I$'s: 
\be\label{quazz}d(M^{IJ}*dB_J)-d(A^I_u dq^u)-\F^\L e^I_\L=0\ee
where we have defined $M^{IJ}h_{JK}=\delta^I_K$ and $h_{Iu}=h_{IJ}A^J_u$.
With respect to the ungauged case, the dualization procedure of the
    scalars $q^I$ gives an electric source term in the equation of
    motion of the $B_I$'s due to the gauge term in the $q^I$'s Bianchi
    identity. Considering as the new equations of motion \eq{quazz}
    and \eq{eomqu} where \eq{sburz} is used, we obtain the new
    Lagrangian: 
\be\mathcal{L}\!=\!\frac{1}{2}\F^\L\Gi_\L\!-\!2e^I_\L\!B_I\F^\L\!\!-
    \!g_{uv}dq^u\!\!*\!dq^v\!\!-\!2A_{u}^Idq^u\!  
    dB_I\!-\!M^{I\!J}\!dB_I\!*\!dB_J\!-\!V\!\!*\!\!\bfone\label{boslageF}\ee 
The gauge terms remain unchanged after dualization \eq{conn}, therefore the
    scalar potential is the same as in equation \eq{pot1} and the
    fermion shifts are given by: 
\ba
&&\nabla\p_A^{(g)}=iS_{AB}\g_a\p^BV^a;\quad
\nabla\l^{iA(g)}=W^{iAB}\p_B;\quad
\nabla\zeta_\a^{(g)}=N^A_\a\p_A\nn\\
&&\label{shift}
S_{AB} = \frac{ i}{2}\sigma^x_{AB} {\Pii}^x_\Lambda L^\Lambda;\quad
W^{iAB}  = i g^{i\bar \jmath}
\sigma^{AB}_x{\Pii}^x_\Lambda\bar f_{\bar \jmath}^\Lambda;\quad 
N^\alpha_A  = -2\mathcal{U}^\alpha_{AI}e^I_\Lambda L^\Lambda
\ea
The reduced quaternionic manifold is described, as already found in
    ref.\cite{Theis:2003jj}, \cite{Dall'Agata:2003yr}, by the new
    vielbeins: 
\be
\label{U}
P^{A\alpha}_{u}\equiv\mathcal{U}_{ u}^{A\alpha}-
A^{I}_u\mathcal{U}_{I}^{A\alpha}
\ee
and the reduced quaternionic connections are related to the old (hatted) ones:
\be
\hat\omega^{AB}_u\!\equiv\omega^{AB}_u+\!A^I_u\omega^{AB}_I;\,
\hat\omega^{AB}_I\equiv\omega^{AB}_I;\   
\hat\Delta^{\alpha\beta}_u\equiv\Delta^{\alpha\beta}_u+
\!A^I_u\Delta^{\alpha\beta}_I;\,
\hat\Delta^{\alpha\beta}_I\equiv\Delta^{\alpha\beta}_I   
\label{conn}
\ee
that is
\ba\hat\o^{AB}&=&\hat\o^{AB}_udq^u+\hat\o^{AB}_I\nabla q^I=
\o^{AB}_udq^u-*H^I\o_I^{AB}\label{cp}\\ 
\mathcal{U}_{\a A}&=&\mathcal{U}_{\a Au}dq^u+ \mathcal{U}_{\a
  AI}\nabla q^I=P_{\a Au}dq^u-\mathcal{U}_{IA\a}*H^I\label{pc} \ea 
Therefore we obtain the following parametrizations for the fermionic fields:
\ba
D\p_A&=&\r_{Aab}V^aV^b+\e_{AB}T^{-}_{ab}\g^b \p^BV^a-*H^I_a\o_{IA}^{\
  \ B}\p_BV^a+iS_{AB}\g_a\p^BV^a\\ 
D \lambda^{iA}&=&D_a \lambda^{iA}V^a + {\rm i}Z^i_a \gamma^a \psi^A +
G^{-i}_{ab}\,\gamma^{ab}\psi_{B}\epsilon^{AB}+W^{iAB}\p_B\\ 
D\zeta_\a&=&D_a\zeta_\a V^a+iP_{aA\a}\g^a\p^A-
i\mathcal{U}_{IA\a}*H^I_a\g^a\p^A+N^A_\a\p_A 
\ea
where the covariant derivatives are defined with the reduced ${\rm
SU}(2)$ and\break ${\rm Sp}(2n,\mathbb{R})$ connections.
Since the closure of the Bianchi identities, once the parametrizations
of the fermions is given, determines the parametrization of the
bosons, the theory we just obtained coincides at the ungauged level
with the one constructed in reference
\cite{Dall'Agata:2003yr}\footnote{We have corrected a misprinted
  factor in the parametrization of the gravitino
  \cite{Dall'Agata:2003yr}.}. The gauge part differs and describes a
``Green--Schwarz'' coupling for the electric fields, as the one
obtained from IIB compactification in presence of electric fluxes,
\cite{Dall'Agata:2001zh} \cite{Louis:2002ny}.  

\section{Magnetic fluxes from dualization}
If we want to consider also the presence of magnetic fluxes we have to
perform a symplectic rotation on the theory. Symplectic
transformations \cite{Gaillard:1981rj}, \cite{Andrianopoli:1996ve} are
invariances of the Bianchi identities/equations of motion for the
ungauged theory but are not in general symmetries of the action as
they may mix  electric and magnetic field strengths. If the gauging is
performed the symmetry between electric and magnetic field strengths
does not hold any more due to the presence of a source term for the
electric field strengths. This is true also for the dual theory with
tensor multiplets as a source term appears just in the vector equation
of motion \eq{bieom}.\\   
If one acts with a symplectic rotation on the Bianchi
identities/equations of motion on the standard gauged supergravity,
one gets also a source for the magnetic field strengths and a
perturbative formulation in terms of a Lagrangian may be
problematic. Nevertheless, a restriction on the symplectic
transformation and  the dualizations of the gauged scalar fields into
tensors allows us to write a Lagrangian for the new theory.\\  
For this purpose, let us formulate the standard $N=2$ theory in a
symplectic covariant setup. We  denote with $<\ ,\ >$ the symplectic
product  
\begin{equation}<a,b>=
\begin{pmatrix}{a^\Lambda &b_\Lambda}\end{pmatrix}
\begin{pmatrix}{0&\delta_\Lambda{}^\Sigma\cr
-{\delta^\Lambda{}_\Sigma}&0}\end{pmatrix}
\begin{pmatrix}{c^\Sigma\cr d_\Sigma}\end{pmatrix}=a^\L d_\L-c^\L b_\L
\end{equation}           
and write the gauge covariant derivative of $q^I$ \eq{decu} as:
\be \nabla q^I=dq^I+<\A,\K^I>=dq^I+\A^{\dot\L}e^I_{\dot\L}\ee
where we have defined  the symplectic vector of electric/magnetic
potentials $\A$ and  the electric/magnetic killing vector $\K^I$ 
\be \A=\begin{pmatrix}{\A^{\dot\L}\cr \A^{\tilde\L}\cr \A_{\dot\L}\cr
    \A_{\tilde\L}}\end{pmatrix};\quad\quad \K^I=\begin{pmatrix}{0\cr
    0\cr e^I_{\dot\L}\cr 0}\end{pmatrix}\ee 
according to:
\be \F^\L=d\A^\L;\quad\quad\Gi_\L=d\A_\L\ee
Note that we would not be allowed to introduce the magnetic potential $\A_{\dot\L}$ for the gauged theory, which is not well defined since $d\Gi_{\dot\L}\neq 0$; nevertheless it does never appear in the Lagrangian  and we will manage to never make it appear. 
We can now perform a symplectic rotation on the theory by means of a matrix $Q$:
\be Q=\begin{pmatrix}{A&B\cr C&D}\end{pmatrix}\label{Q}\ee
where, for the moment, $A$, $B$, $C$, $D$ are taken to be generic matrixes which satisfy the relations which define a symplectic matrix:
\be A^TC=C^TA;\quad B^TD=D^TB;\quad A^TD-C^TB=\bfone\label{symcon}
\ee
>From the rotation of the vectors $\A$, $\K^I$ we obtain:
\ba\label{rotvect} \A^\prime&=&\begin{pmatrix}{\A^{\dot\L\,\prime}\cr \A^{\tilde\L\,\prime}\cr \A_{\dot\L}^{\prime}\cr \A_{\tilde\L}^{\prime}}\end{pmatrix}=\begin{pmatrix}
{A^{\dot\L}_{\dot\S}\A^{\dot\S}+A^{\dot\L}_{\tilde\S}\A^{\tilde\S}+B^{\dot\L\dot\S}\A_{\dot\S}+B^{\dot\L\tilde\S}\A_{\tilde\S}\cr
A^{\tilde\L}_{\dot\S}\A^{\dot\S}+A^{\tilde\L}_{\tilde\S}\A^{\tilde\S}+B^{\tilde\L\dot\S}\A_{\dot\S}+B^{\tilde\L\tilde\S}\A_{\tilde\S}\cr
C_{\dot\L\dot\S}\A^{\dot\S}+C_{\dot\L\tilde\S}\A^{\tilde\S}+D_{\dot\L}^{\dot\S}\A_{\dot\S}+D_{\dot\L}^{\tilde\S}\A_{\tilde\S}\cr
C_{\tilde\L\dot\S}\A^{\dot\S}+C_{\tilde\L\tilde\S}\A^{\tilde\S}+D_{\tilde\L}^{\dot\S}\A_{\dot\S}+D_{\tilde\L}^{\tilde\S}\A_{\tilde\S}}
\end{pmatrix}\\ \nn\\
\label{rotkill}\K^{I\,\prime}&=&\begin{pmatrix}{m^{I\dot\L\,\prime}\cr m^{I\tilde\L\,\prime}\cr e^{I\,\prime}_{\dot\L}\cr e^{I\,\prime}_{\tilde\L}}\end{pmatrix}=\begin{pmatrix}{B^{\dot\L\dot\S}e^I_{\dot\S}\cr B^{\tilde\L\dot\S}e^I_{\dot\S}\cr D_{\dot\L}^{\dot\S} e^I_{\dot\S}\cr D_{\tilde\L}^{\dot\S}e^I_{\dot\S} }\end{pmatrix}
\ea
If we define the symplectic vector of the new electric and magnetic field strengths as $\mathcal{H}^\prime=d\A^\prime$
the new equations of motion/Bianchi identities are:
\be\label{rotbieom}d\mathcal{H}^\prime=\K^{I\,\prime}J_I\ee
The covariant derivative of the scalar becomes:
\be\label{covdernew}\nabla q^I=dq^I+\A^{\dot\L\,\prime}e^{I\,\prime}_{\dot\L}+\A^{\tilde\L\,\prime}e^{I\,\prime}_{\tilde\L}-\A_{\dot\L}^\prime m^{I\dot\L\,\prime}-\A_{\tilde\L}^\prime m^{I\tilde\L\,\prime}\ee
We easily see from equation \eq{covdernew}, \eq{rotvect} that for a generic symplectic rotation the non well defined magnetic potential $\A_{\dot\L}$ would appear in the action explicitly, therefore we have to restrict the matrix $Q$ \eq{Q} to the transformations which avoid the presence of  $\A_{\dot\L}$ in the covariant derivative of the scalar fields $\nabla q^I$ \eq{covdernew}. This restriction holds also if we want to dualize these scalars into tensors, since as we can see from equation \eq{sburz}, if $\nabla q^I$ is not well defined also the field strength $H_I=dB_I$ would not be well defined.\\  
In order to obtain a sensible but non trivial result, we can restrict the transformation matrix $Q$ to have some vanishing blocks, for instance: 
\be\label{restrict}B^{\dot\L\dot\S}=D_{\tilde\L}^{\,\dot\S}=0\ee  
This is enough to obtain:
\be m^{I\dot\L\,\prime}=e^{I\prime}_{\tilde\L}=0\label{ort} \ee
and therefore:
\be\label{nabq} \K^{I\,\prime}=\begin{pmatrix}{0\cr m^{I\tilde\L\,\prime}\cr e^{I\prime}_{\dot\L}\cr 0}\end{pmatrix}\ \Rightarrow\ \nabla q^I=dq^I+\A^{\dot\L\,\prime}e^{I\,\prime}_{\dot\L}-\A_{\tilde\L}^\prime m^{I\tilde\L\,\prime}\ee 
Note from equation \eq{rotvect} that thanks to equation \eq{restrict}, the ill defined magnetic potential $\A_{\dot\S}$ does not enter the definition of the new potentials $\A^{\dot\L\,\prime}$ and $\A_{\tilde\L}^{\prime}$ which appear in the definition \eq{nabq}. In this case equation\eq{rotbieom} becomes:
\be\label{newbieom} d\begin{pmatrix}{\F^{\dot\L\,\prime}\cr\Gi_{\dot\L}^\prime}\end{pmatrix}=\begin{pmatrix}{0\cr e^{I\,\prime}_{\dot\L}J_I}\end{pmatrix};\quad d\begin{pmatrix}{\F^{\tilde\L\,\prime}\cr\Gi_{\tilde\L}^\prime}\end{pmatrix}=\begin{pmatrix}{ m^{I\tilde\L\,\prime}J_I\cr 0}\end{pmatrix}\ee
Let us now turn to the problem of writing an action which gives equations \eq{newbieom}; one immediately realizes that the problem is the presence of the magnetic current, nevertheless we recall that the equation of motion for the fields $q^I$ \eq{eomqI} corresponds to $dJ_I=0$ which we can solve setting:
\be\label{solvJ} J_I=2dB_I\ \Longrightarrow\ dJ_I=2d^2B_I=0\ee
>From equation \eq{newbieom}, using the condition \eq{restrict} and equation \eq{solvJ} we can read the definitions of $\F^{\dot\L\,\prime}$, $\F^{\tilde\L\,\prime}$, $ \Gi_{\dot\L}^{\prime}$ and  $ \Gi_{\tilde\L}^{\prime}$. In fact,defining:
\ba
\Ci^{\tilde\L}&=&A^{\dot\L}_{\dot\S}\A^{\dot\S}+A^{\dot\L}_{\tilde\S}\A^{\tilde\S}+B^{\dot\L\tilde\S}\A_{\tilde\S};\,\ \quad \Ci_{\tilde\L}=\A^\prime_{\tilde\L}\\
\Ci_{\dot\L}&=&C_{\dot\L\dot\S}\A^{\dot\S}+C_{\dot\L\tilde\S}\A^{\tilde\S}+D_{\dot\L}^{\tilde\S}\A_{\tilde\S};\quad\Ci^{\dot\L}=\A^{\dot\L\,\prime}
\ea
we obtain the expressions for the new electric/magnetic field strengths:
\ba\label{deffs}
\F^{\dot\L\,\prime}&=&d\Ci^{\dot\L}\quad\quad\ \,\quad\quad\quad\quad\quad \F^{\tilde\L\,\prime}=d\Ci^{\tilde\L\,\prime}+2m^{I\tilde\L\,\prime}B_I\nn\\
\Gi_{\dot\L}^{\prime}&=&d\Ci_{\dot\L}+2e^{I\,\prime}_{\dot\L}B_I\quad\quad\quad\quad \Gi^{\prime}_{\tilde\L}=d\Ci_{\tilde\L}
\ea
which obviously fulfill equations \eq{newbieom}, \eq{solvJ}. Note, that defining $\F^{\tilde\L\,\prime},\,\Gi^\prime_{\dot\L}$ in equation \eq{deffs} we integrated equation \eq{newbieom} using equation \eq{solvJ}
\be d\Gi_{\dot\L}=2e^I_{\dot\L}dB_I\quad\longrightarrow\quad \Gi_{\dot\L}=2e^I_{\dot\L}(B_I+d\L_I)\ee
therefore $B_I$ is defined up to an exact 2--form
\be B_I\rightarrow B_I+d\L_I\ee
where $\L_I$ is a generic 1--form. In order to maintain the invariance of the action one must simultaneously redefine
\be \Ci^{\tilde\L}\rightarrow \Ci^{\tilde\L}-2m^{I\tilde\L\,\prime}\L_I \ee  
In order to determine completely the Lagrangian which describes the resulting theory, we note that setting $J_I=2dB_I$ means that we are dualizing the scalars $q^I$ into the tensors $B_I$. Therefore we derive the equation of motion for the fields $B_I$ from the Bianchi identity for the fields $q^I$ as before \eq{sburz}, where now the covariant derivative of the $q^I$'s is given by equation \eq{nabq} and we obtain the equation of motion for the tensor fields $B_I$:
\be\label{eomB}d(M^{IJ}*dB_J)+d(A^I_u dq^u)-(\F^{\L\,\prime}e^{I\,\prime}_\L-\Gi^\prime_\L m^{I\L\,\prime})=0\ee
As before we can write the dualized Lagrangian, which has the following form:
\be\mathcal{L}\!=\!\frac{1}{2}\F^{\L\prime}\!\Gi_\L^\prime\!\!-\!\!2e^{I\prime}_\L\!B_I\F^{\L\prime}\!\!\!-\!g_{uv}dq^u\!\!*\!dq^v\!-\!2A_{u}^Idq^u\!dB_I\!-\!\!M^{I\!J}\!dB_I\!*\!dB_J\!-\!V^\prime\!\!*\!\!\bfone\label{boslagmGeF}\ee
Some comments are in order for the rotated potential $V^{\prime}$. We
can suppose that the symplectic covariant form for the potential $V$
has the following structure: 
\be\label{assumpt}
V=-\frac{1}{2}\begin{pmatrix}{\Q^{x\L}&\Pii^x_\L}\end{pmatrix}\begin{pmatrix}{\mathcal{M}_{\L\S}&\mathcal{M}_{\L}^{\
      \S}\cr\mathcal{M}^{\L}_{\
      \S}&\mathcal{M}^{\L\S}}\end{pmatrix}\begin{pmatrix}{\Q^{x\S}\cr
    \Pii^x_\S}\end{pmatrix}\ee where the matrix $\mathcal{M}$ which contracts the vectors of electric/magnetic prepotentials has to be symmetric. We know that
\be\mathcal{M}^{\L\S}=(Im\N^{-1})^{\L\S} \ee
since the potential \eq{pot1} corresponds to the particular case
$\Q^{x\L}\equiv m^{I\L} \o_I^x$. 
Furthermore as the fermion shifts
turn out to be symplectic invariant quantities
\eq{shiftSr}--\eq{shiftNr}, thanks to the Ward identity
\eq{ward}  we
also know that the scalar potential \eq{assumpt} must be
symplectic invariant.  
Therefore, given one block of such a symmetric symplectic invariant
matrix \eq{assumpt}
one can determine, with suitable symplectic rotations, all the
remaining submatrices. $\mathcal{M}$ turns out to be:
\be \mathcal{M}=\begin{pmatrix}{(I+R I^{-1}R)_{\L\S}&-(R I^{-1})_{\L}^{\ \S}\cr -( I^{-1}R)^{\L}_{\ \S}&(I^{-1})^{\L\S}}\end{pmatrix}\ee 
that is invariant under symplectic transformations, which act both linearly on the matrix and non linearly on the entries (we indicate $I=Im\N$, $R=Re\N$).\\
The transformed potential is therefore given by:
\be V^\prime=-\frac{1}{2}\begin{pmatrix}{\Q^{x\L}&\Pii^x_\L}\end{pmatrix}\begin{pmatrix}{(I+R I^{-1}R)_{\L\S}&-(R I^{-1})_{\L}^{\ \S}\cr -( I^{-1}R)^{\L}_{\ \S}&(I^{-1})^{\L\S}}\end{pmatrix}\begin{pmatrix}{\Q^{x\S}\cr \Pii^x_\S}\end{pmatrix}\label{rotpot} \ee
After the symplectic rotation, the fermion shifts \eq{shift} are modified and become:
\begin{eqnarray}
S_{AB}^\prime & =& \frac{ i}{2}\sigma^x_{AB}\left({\Pii}^{x\,\prime}_\Lambda L^{\Lambda\,\prime} -\Q^{x\L\,\prime} M_\Lambda^\prime\right)= \frac{ i}{2}\sigma^x_{AB}\o^x_I<V,\K^I>^\prime
\label{shiftSr}\\
W^{iAB\,\prime} & = & ig^{i\bar \jmath}
\sigma^{AB}_x\left(\Pii^{x\,\prime}_\Lambda\bar f_{\bar \jmath}^{\Lambda\,\prime}
- \Q^{x\Lambda\,\prime} h_{\Lambda \bar \jmath}^\prime\right)= ig^{i\bar \jmath}\sigma^{AB}_x \o^x_I<U_{\overline{\jmath}},\K^I>^\prime
\label{shiftWr}\\
N^{\alpha\,\prime}_A & = &-2\mathcal{U}^\alpha_{AI}
\left(e^{I\,\prime}_\Lambda L^{\Lambda\,\prime}-
  m^{I\Lambda\,\prime}M_{\Lambda}^\prime\right)
=-2\mathcal{U}^\alpha_{AI}<V,\K^I>^\prime 
\label{shiftNr}
\end{eqnarray}
where the restriction \eq{ort} holds. The supersymmetry ward identity
\be\d^A_B V=-12S^{AC}S_{CB}+g_{i\overline\jmath}W^{iAC}
W^{\overline\jmath}_{CB}+2N^A_\a N^\a_B\label{ward}\ee 
imposes the constraint
\be m^{\L[I} e^{J]}_\L = \mez <\K^I,\K^J>=0\label{tadpole} \ee 
which in this case it is satisfied thanks to \eq{ort}.\\
We also checked the assumption \eq{assumpt} on the structure of the
scalar potential is correct, computing explicitly the scalar potential
\eq{rotpot} using the fermion shifts \eq{shiftSr}--\eq{shiftNr} into
equation \eq{ward}.\\ 
In the dualization procedure discussed above there are two important points.\\
The first is that the symplectic rotation which one has to perform in
order to generate the ``magnetic'' part of the symplectic product
\eq{shiftSr}--\eq{shiftNr} has to be done with a matrix whith $B\neq
0$, which corresponds to a non perturbative transformation which
relates a perturbative gauge theory with coupling constant $g$ with a
perturbative theory with coupling constant $1/g$ as
$g_{\L\S}=-Im\N_{\L\S}$.\\ 
Conversely a supergravity theory with non vanishing $m^{I\L}$ is not
equivalent to a standard supergravity with translational gauging, as
one has to perform a symplectic transformation which is not an
invariance of the action.\\ 
The second is that if we perform such a symplectic rotation on the
standard gauged supergravity we are forced to dualize the scalars
associated to the translational isometries into tensors in order to be
able to write an action, which means that the symplectic extension of
the standard $N=2$ theory \cite{Andrianopoli:1996cm} can be formulated
just in terms of scalar--tensor multiplets
\cite{Michelson:1996pn}. Further evidences of this fact can be
provided if one imposes the closure of the Bianchi identities as it
was done in reference \cite{Dall'Agata:2003yr}, and will be specified
to the present case in a forthcoming paper \cite{nuovo}.\\ 
If we want to make contact with the theory obtained from type IIB
compactification on Calabi--Yau in presence of fluxes we consider a
double tensor multiplet $(B_I,\t)$, $I=1,2$. The $B_I$'s descend from
the RR and the NSNS 2--forms while $\t=l+ie^{-\phi}$ is the ten
dimensional complex dilaton. We find \cite{Ferrara:ik} that the
connections $\o_I^x$ are given by: 
\be
 \o_1^{(1)}\!=0;\ \o_1^{(2)}\!=0;\ \o_1^{(3)}\!=e^\varphi;\
 \o_2^{(1)}\!=-e^\varphi Im\t;\
 \o_2^{(2)}\!=0;\,\o_2^{(3)}\!=e^\varphi Re\t\label{oconn} 
\ee
According to the definition \eq{oconn} the scalar potential becomes:
\ba
V&=&-\frac{1}{2}e^{2\varphi}\left[\left(e_\L-\overline{\N}_{\L\S}m^\S
  \right)\left(Im\N^{-1}\right)^{\L\G}\left(\overline{e}_\G-\N_{\G\D}
    \overline{m}^\D\right)\right]+\nn\\[2mm]&&+2
Im\t\left(e_{1\L}m_2^\L-e_{2\L}m_1^\L\right)\ea   
where
\be e_\L\equiv e^1_\L+\t e^2_\L;\quad\quad m^\L\equiv m^{1\L}+\t m^{2\L}\ee
which coincides with the potential of reference \cite{Taylor:1999ii}
once it is written in the supergravity frame. The tadpole cancellation
\eq{tadpole} is now imposed by the supersymmetry Ward identity
\eq{ward}. 
\vskip 1cm
\noindent
\newpage
{\bf Acknowledgments}
\vskip 0.5cm
\noindent We would like to thank R. D'Auria, G. Dall'Agata,
M. Trigiante and H. Samtleben for valuable discussions. L. S. work is
supported in part by the European Community's Human Potential
Programme under contract HPRN--CT--2000--00131 Quantum Spacetime, in
which he is associated to Torino University.

\end{document}